\documentclass[preprint,12pt]{elsarticle}

\usepackage{graphicx}

\usepackage{amssymb}
\usepackage{amsthm}
\usepackage{bm}
\usepackage{upgreek}
\usepackage{subfigure}
\usepackage{multirow}
\usepackage{booktabs}
\usepackage{caption}
\usepackage{url}
\usepackage{epsfig}

\usepackage{lineno}

\journal{Computer Physics Communication}

\begin{document}

\begin{frontmatter}



%
\title{Kalman-filter-based track fitting in non-uniform magnetic field with segment-wise helical track model}
%

\author[tuhep]{Bo Li}
\author[kek]{Keisuke Fujii\corref{corauthor}}\ead{keisuke.fujii@kek.jp}
\author[tuhep]{Yuanning Gao}

\cortext[corauthor]{Corresponding author.}

\address[tuhep]{Department of Engineering Physics, Tsinghua University, Beijing, 100084, China}
\address[kek]{High Energy Accelerator Research Organization (KEK), Tsukuba, 305-0801, Japan}

\begin{abstract}
In the future International Linear Collider (ILC) experiment, high performance tracking is essential to its physics program including precision Higgs studies.
One of major challenges for a detector such as the proposed International Large Detector (ILD) is to provide excellent momentum resolution in a magnetic filed with small (but non-negligible) non-uniformity.
The non-uniform magnetic field implies deviation from a helical track and hence requires the extension of a helical track model used for track fitting in a uniform magnetic field.
In this paper, a segment-wise helical track model is introduced as such an extension. 
The segment-wise helical track model approximates the magnetic field between two nearby measurement sites to be uniform and steps between the two sites along a helix. The helix frame is then transformed according to the new magnetic field direction for the next step, so as to take into account the non-uniformity of the magnetic field.
Details of the algorithm and mathematical aspects of the segment-wise helical track model in a Kalman-filter-based track fitting in the non-uniform magnetic field are elaborated.
The new track model is implemented and successfully tested in the framework of the Kalman filter tracking software package, KalTest, which was originally developed for tracking in a uniform magnetic field.
\end{abstract}


\end{frontmatter}

\linenumbers


\section{Introduction}\label{sec::introduction}
One of the primary goals of the next generation $e^{+} e^{-}$ collider, such as the International Linear Collider (ILC), is to make precise measurements of the properties of the Higgs boson thereby uncovering the secret of the electroweak symmetry breaking. This physics goal imposes great challenges on ILC detectors. For a main tracking detector, for instance, the momentum resolution $\delta(1/p_{\mathrm{T}})$ is required to be $O(10^{-4})\ (\mathrm{GeV/}c)^{-1}$ or better. The International Large Detector (ILD), one of the two conceptual detector designs currently pursued for the ILC experiments, uses a Time Projection Chamber (TPC) as its main tracking detector\cite{Abe:2010aa}. For the ILD TPC, the above required momentum resolution translates into about 200 sampling points along a track with a transverse spatial resolution of $100\ \upmu\mathrm{m}$ or better over its full drift length of $2.2\ \mathrm{m}$ in a magnetic field of $3.5\ \mathrm{T}$. New TPC readout techniques, based on Micro Pattern Gaseous Detector (MPGD) technologies having small $\boldsymbol{E}\times\boldsymbol{B}$ effect, good two-hit resolution, and excellent spatial resolution, provide a promising solution to satisfy the rigorous demand of the ILC.

In order to fulfill the required performance, however, the hardware R\&Ds have to be backed  by  software developments that match the environment of the linear collider TPC.
A Kalman fitler software package, KalTest, has been successfully used for tracking in full detector simulations for physics feasibility studies as well as for tracking in test beam data taken with a Large Prototype (LP) TPC\cite{Abernathy:2010zz}. For the LP test, where the effect of the non-uniform magnetic field on the particle trajectory is small\footnote{
There are two major ways for the presence of a non-uniform magnetic field to manifest itself:
(1) $\boldsymbol{E}\times\boldsymbol{B}$ effect that distorts the measured hit points and hence the apparent trajectory and (2) the real deviation of a track from a helical trajectory. The former effect can be corrected away in principle. As long as the latter is negligible, we can therefore use the helical track model.
}, we could use a helical track model as used in the original KalTest. However, the future real LCTPC such as the ILD TPC must work in a non-uniform magnetic field with a non-uniformity up to a few percent\cite{Peter:2010}. In principle, Kalman filter algorithm itself is independent of the track model and can adapt to the non-uniform magnetic field. The most general solution is to implement a generic track model together with a Runge-Kutta track propagator. For the modest non-uniformity, this solution might not be optimal from the CPU time point of view. In this paper, another solution, a segment-wise helical track model, is proposed. It will be shown that the implementation of the segment-wise helical track model that replaces the original simple helical track model allows us to successfully realize Kalman-filter-based track fitting in a moderate non-uniform magnetic field with the minimal change to the original KalTest package. The time consumption of track fitting in the non-uniform magnetic field will also be discussed.

\section{KalTest}\label{sec::kaltest}
KalTest is a ROOT\cite{Antcheva:2011zz} based Kalman filter software package written in C++ for track fitting in high energy physics experiments. Comparing with the least square fitting method, the Kalman filter has great advantages in track fitting\cite{Fruhwirth:1987fm}. The basic formulae and their implementation in KalTest are summarized in this section, while details can be found in KalTest manual\cite{Keisuke:Manual}. 
\subsection{Kalman filter}

The Kalman filter handles a system that evolves according to an equation of motion ({\it system equation}) under the influence of random disturbance ({\it process noise}). 
It is designed to provide the optimal estimate of the system's state at a given point from the information collected at multiple observation points ({\it measurement sites}). 
Suppose that there are $n$ measurement sites ($k=1,\cdots,n$)
and the state of the system at site ($k$) can be specified by a $p$-dimensional column vector ({\it state vector}) $\boldsymbol{\bar a}_{k}$, where the bar indicates that it is the true state vector without any measurement error. 
The system equation that describes the evolution of the state at site ($k-1$) to the next one, site ($k$), can be written in the form:
\begin{equation}
\boldsymbol{\bar a}_{k} = 
	\boldsymbol{f}_{k-1} ( \boldsymbol{\bar a}_{k-1} )  + \boldsymbol{w}_{k-1},
\label{EQ:ak}
\end{equation}
where $\boldsymbol{f}_{k-1} (\boldsymbol{\bar a}_{k-1})$ is a {\it state propagator} which expresses a smooth and deterministic motion that would take place if there were no process noise, and $\boldsymbol{w}_{k-1}$ is the  {\it process noise} term due to the random disturbance. 
It is assumed that the process noise is unbiased and has a covariance given by
\begin{equation}
\boldsymbol{Q}_{k-1} \equiv  \,
	< \boldsymbol{w}_{k-1} \boldsymbol{w}_{k-1}^{T} > .
\label{EQ:Qk-1}
\end{equation}

At each site, we measure some observables about the system.
The values of these observables comprise a $m$-dimensional column vector ({\it measurement vector}) 
$\boldsymbol{m}_{k}$.
Its relation to the state vector $\boldsymbol{\bar a}_{k}$ at site ($k$) is called a {\it measurement equation}:
\begin{equation}
\boldsymbol{m}_{k} = 
	\boldsymbol{h}_{k}( \boldsymbol{\bar a}_{k}) + \boldsymbol{\epsilon}_{k},
\label{EQ:mk}
\end{equation}
in which $\boldsymbol{h}_{k}(\boldsymbol{\bar a}_{k})$ is a {\it projector} which gives, as a function of the state vector, the measurement vector you would expect for an ideal measurement with no measurement error, and $\boldsymbol{\epsilon}_{k}$ is the random measurement error ({\it measurement noise}) unavoidable in practice.
We assume here that systematic errors such as those from misalignment of detectors have been corrected and hence  the random measurement noise is unbiased and having a covariance given by
\begin{equation}
\boldsymbol{V}_{k} \equiv  (\boldsymbol{G}_k)^{-1} \equiv \,
	<  \boldsymbol{\epsilon}_{k}  \boldsymbol{\epsilon}_{k}^{T}> .
\label{EQ:Vk}
\end{equation}

In the Kalman filter process, two operations, {\it prediction} and {\it filtering}, are needed at each site to proceed.
The state vector prediction is the extrapolation of $\boldsymbol{a}_{k-1}^{k-1}$
to the next site by using Eq.(\ref{EQ:ak}):
\begin{equation}
\boldsymbol{a}_{k}^{k-1} 
	= \boldsymbol{f}_{k-1}  (  \boldsymbol{a}_{k-1}^{k-1} ) \equiv  \boldsymbol{f}_{k-1}  (  \boldsymbol{a}_{k-1} ),
\label{EQ:akk-1}
\end{equation}
where the superscripts ($k-1$) to the state vectors indicate that the state vectors are estimated using the information up to site ($k-1$). In what follows we will omit the superscript, if the superscript coincides with the subscript as $\boldsymbol{a}_{k-1} \equiv \boldsymbol{a}_{k-1}^{k-1}$.

The covariance matrix for $\boldsymbol{a}_{k-1}$ is defined by
\begin{equation}
\boldsymbol{C}_{k-1} 
 \equiv  
  \left< \left(\boldsymbol{a}_{k-1} - \boldsymbol{\bar a}_{k-1}\right)
 \left(\boldsymbol{a}_{k-1} - \boldsymbol{\bar a}_{k-1}\right)^T\right> ,
\label{EQ:Ck-1}
\end{equation}
then the prediction for the covariance matrix at site ($k$) is given by
\begin{equation}
\label{EQ:Ckk-1}
\boldsymbol{C}_{k}^{k-1}  = \boldsymbol{F}_{k-1} \boldsymbol{C}_{k-1} \boldsymbol{F}_{k-1}^T + \boldsymbol{Q}_{k-1},
\end{equation}
where
\begin{equation}
\boldsymbol{F}_{k-1} \equiv  \frac{\partial \boldsymbol{f}_{k-1}}{\partial \boldsymbol{a}_{k-1}}
\end{equation}
is called a {\it propagation matrix}.

In the filtering step, the predicted state vector at site ($k$) is updated by taking into account the pull that is defined to be the difference between the measured and the predicted measurement vectors, $\boldsymbol{m}_k-\boldsymbol{h}_k (\boldsymbol{a}_k^{k-1})$, as
\begin{equation}\label{eq::filter}
\boldsymbol{a}_k=
\boldsymbol{a}_k^{k-1}+
\boldsymbol{K}_k
\left(
\boldsymbol{m}_k-\boldsymbol{h}_k (\boldsymbol{a}_k^{k-1})
\right),
\end{equation}
in which, $\boldsymbol{K}_k$ is the gain matrix given by
\begin{equation}\label{eq::gain}
  \boldsymbol{K}_{k}  =  
       \left[ \left( \boldsymbol{C}_{k}^{k-1}\right)^{-1} 
                    + \boldsymbol{H}_{k}^T \boldsymbol{G}_{k} \boldsymbol{H}_{k} \right]^{-1} 
       \boldsymbol{H}_{k}^T \boldsymbol{G}_{k} 
\end{equation}
with $\boldsymbol{H}_{k}$ defined by
\begin{equation}\label{EQ:Hk} 
\boldsymbol{H}_{k}  \equiv
  \frac{\partial \boldsymbol{h}_{k}}{\partial \boldsymbol{a}_{k}^{k-1}} ,
\end{equation}
which is called the {\it measurement matrix}.

After all the $n$ sites are filtered, the state vector at site ($k(k<n)$) can be reevaluated by including the information at subsequent sites: $k+1$ to $n$. This process is called {\it Smoothing}.
The smoothed state at site ($k$) is obtained by the following backward recurrence formula
\begin{equation}
\left\{
\begin{array}{lcl}
\boldsymbol{a}_{k}^{n} & = & \boldsymbol{a}_{k} 
        + \boldsymbol{A}_{k} ( \boldsymbol{a}_{k+1}^{n} - \boldsymbol{a}_{k+1}^{k} )
\cr
\boldsymbol{A}_{k} & = & 
   \boldsymbol{C}_{k} \boldsymbol{F}_{k}^T \left( \boldsymbol{C}_{k+1}^{k} \right)^{-1} 
\end{array}
\right. ,
\label{EQ:Ak}
\end{equation}
which gives the smoothed state at site ($k$) in terms of the smoothed state at site ($k+1$), the predicted state at site ($k+1$), and the filtered state at site ($k$).

\subsection{Helical track parametrization}
\begin{figure}[!htb]
\centering
\includegraphics[scale=0.6]{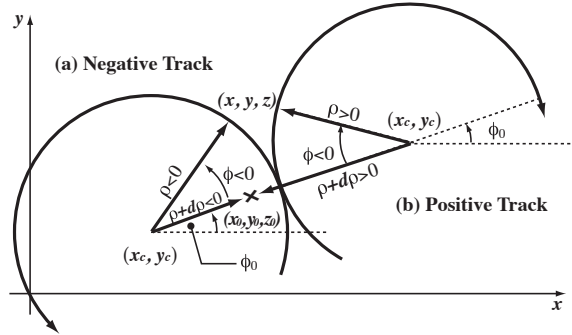}
\caption{Helical track parametrization in KalTest. $\boldsymbol{x}=(x, y, z)$ is a point on the helix, $\boldsymbol{x}_{0}=(x_0, y_0, z_0)$ a reference point usually taken to be a hit point. $d_{\rho}$ is the distance of the helix from the pivot in the $x \textrm{-} y$ plane, $\phi_{0}$ the azimuthal angle of the pivot with respect to the center of the helix,  $d_{z}$ the distance of the helix from the pivot in the $z$ direction, and $\lambda$ stands for the dip angle. }
\label{fig:helix}
\end{figure}

In a uniform magnetic field a charged particle follows a helical trajectory. 
If we set our coordinate system in such a way that the magnetic field points to the $z$ axis direction, the helix can be parametrized as
\begin{equation}\label{eq:helixequation}
\left\{
\begin{array}{lllllllll}
x&=&x_0&+&d_{\rho}\cos\phi_{0}&+&\frac{\alpha}{\kappa}
(\cos\phi_{0}-\cos(\phi_{0}+\phi))\\
y&=&y_0&+&d_{\rho}\sin\phi_{0}&+&\frac{\alpha}{\kappa}
(\sin\phi_{0}-\sin(\phi_{0}+\phi))\\
z&=&z_{0}&+&d_{z}&-&\frac{\alpha}{\kappa}\tan\lambda \cdot \phi
\end{array}, 
\right.
\end{equation}
where $\boldsymbol{x}_{0}=(x_0, y_0, z_0)$ is an arbitrary reference point 
on which three ($d_\rho$, $\phi_0$, and $d_z$) out of the five helix parameters, $d_\rho$, $\phi_0$, $\kappa$, $d_z$, and $\tan\lambda$, depend 
and $\alpha$ is a constant defined by
$\alpha \equiv 1/cB$ with $B$ and $c$ being the magnetic field and the speed of light, respectively.
Since the reference point $\boldsymbol{x}_0$ is arbitrary, we can take it to be the measured hit point at each site, say, site ($k$) and call it a pivot.
Then $\phi$ measures the deflection angle from the pivot.
The geometrical meanings of the five helix parameters are depicted in Fig.\ref{fig:helix}.
Notice that $\rho \equiv \alpha/\kappa$ is the radius of the helix singed by the particle charge, while
$\kappa \equiv Q/p_t$ with $Q$ being the charge in units of the elementary charge and $p_t$ being the transverse momentum.
%
%
The five parameters in Eq.(\ref{eq:helixequation}) are combined to make a concrete state vector 
\begin{displaymath}
\boldsymbol{a}_{k}=
\left(
\begin{array}{c}
d_{\rho},
\phi_{0},
\kappa,
d_{z},
\tan\lambda
\end{array}
\right)^\mathrm{T}.
\end{displaymath}
The reason to use $\kappa = Q/p_t$ instead of $\rho$ or $p_t$ as a helix parameter is to allow continuous change of the state vector for a high momentum track for which the curvature might change its sign during the track fitting.
Notice that there is a difference of $\pi$ in the definition of $\phi_0$, depending on the track charge. This is to avoid discontinuity of $\phi_0$ that would happen if $\kappa$ changes its sign during track fitting.

\subsection{Implementation}
Notice that the formalism presented above is generic, though the concrete forms of the state propagator $\boldsymbol{f}_{k-1}(\boldsymbol{a}_{k-1})$, the projector $\boldsymbol{h_{k}}(\boldsymbol{a}_{k}^{k-1})$, and their derivatives, $\boldsymbol{F}_{k-1}$ and $\boldsymbol{H}_{k}$, depend on the track model and the geometry and distribution of the measurement sites.
In order to keep the application scope of this generic formalism as wide as possible, KalTest is designed to extract this generic part as a separate sub-package (KalLib) that forms an abstract layer consisting of abstract base classes that implement the generic Kalman filter algorithm.
In order to realize Kalman-filter-based track fitting, we then inherit from the generic base classes in the abstract base class library KalLib and implement their pure virtual methods for track fitting in a tracking detector consisting of measurement layers.
The resultant  Kalman-filter-based track fitter library is KalTrackLib. 
KalTrackLib, however, should not depend on any particular track model or shape or coordinate system of any measurement layer so as to accommodate variety of tracking devices which may coexist in the same tracking system.
KalTrackLib is hence designed to interact with concrete track model or concrete measurement layer classes always through interface classes that represent an abstract track or an abstract measurement layer class.
The implementation of concrete track models (helix, straight line, etc.) and measurement surfaces (cylinder, hyperbolic, flat plane, conical surface, etc.) are separated out into a geometry library eomLib.

\begin{figure}[!htb]
\centering
\includegraphics[scale=0.7]{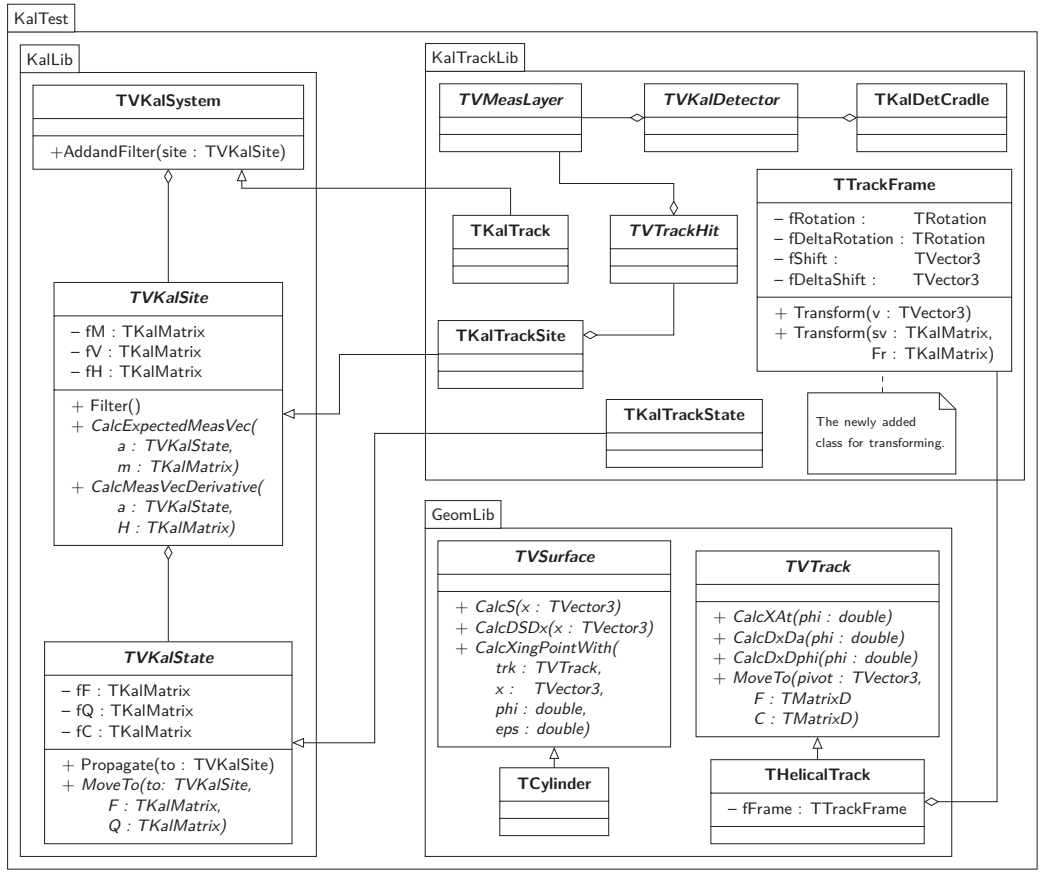}
\caption{Class diagram of KalTest. KalLib contains the generic Kalman filter procedure. The pure virtual functions specific to tracking are implemented in KalTrackLib. Basic geometrical objects to express detector configuration or to represent track models are implemented in GeomLib.}
\label{fig::classdiag}
\end{figure}

Figure \ref{fig::classdiag} shows some major classes contained in these three packages. 
In the KalLib library, {\tt TVKalSystem} is implemented as an array of {\tt TVKalSite}-derived objects and represents a collection of information gathered at each measurement site.
Each {\tt TVKalSite}-derived object can contain up to  three Kalman filter states (i.e. predicted, filtered, and smoothed), each of them being an instance of some concrete class inheriting from {\tt TVKalState}.
The functions {\tt TVKalState::Propagate()} and {\tt TVKalSite::Filter()} implement Eqs.(\ref{EQ:Ckk-1}) and (\ref{eq::filter}), respectively.
The pure virtual functions such as {\tt TVKalState::MoveTo()},  {\tt TVKalSite::CalcExpectedMeasVec()}, and {\tt TVKalSite::CalcMeasVecDerivative()} declared in the two classes are implemented in the corresponding derived classes of KalTrackLib. 
These concrete functions in the derived classes interact with a track model or a measurement layer through the abstract layer made of {\tt TVTrack}, {\tt TVSurface}, and {\tt TVMeasLayer} and are still quite generic.
The concrete geometrical features of the track or the measurement layers are supplied by concrete classes in GeomLib, where, for instance, the helical track model defined by Eq.(\ref{eq:helixequation}) is implemented in a class called {\tt THelicalTrack} derived from {\tt TVTrack}, which actually propagates and calculates the concrete value of the state vector $\boldsymbol{a}$, the propagation matrix $\boldsymbol{F}$, and the covariance matrix $\boldsymbol{C}$ and a cylindrical surface is implemented in a class called {\tt TCylinder} inheriting from {\tt TVSurface}.

The architectural design of KalTest thus minimizes the number of user-implemented classes to the following three:
\begin{itemize}
\item {\bf MeasLayer}: a measurement layer object that multiply inherits from a concrete shape class such as {\tt TCylinder}  and the abstract measurement layer class {\tt TVMeasLayer}.
\item {\bf KalDetector}: a class derived from {\tt TVKalDetector}, which is implemented as an array of {\tt TVMeasLayer} pointers and holds MeasLayers with any shape and coordinate system as well as materials.
\item {\bf Hit}: a coordinate vector class as defined by the MeasLayer class, which inherits from {\tt TVTrackHit}.
\end{itemize}
Notice that, by design, KalTest allows site-to-site change of track models.

Since the original helical track model in KalTest is valid only in a uniform magnetic field, the result of track fitting will not be satisfactory if the track is generated in a significantly non-uniform magnetic field, as we will see in Sec.\ref{sec::results}. 
A new class in the KalTrackLib library, {\tt TTrackFrame}, which contains the algorithm for transforming a coordinate frame in a non-uniform magnetic field, can solve this issue and will be described in the next section. 

\section{Algorithm for non-uniform magnetic field}\label{sec::algorithm}
\subsection{Basic idea}\label{subsec::basicidea}
If the non-uniformity of the magnetic field is not too large, we can assume that the magnetic field between two measurement layers is approximately uniform.
We can then propagate our track from one measurement layer to the next using the helix model we discussed above.
When the track reaches the next layer, we update the magnetic field in order to take the non-uniformity into account,
and propagate the track with the updated magnetic field.
The track produced this way is hence segment-wise helical and hereafter called a {\it segment-wise helical track}.
Notice that the direction of the magnetic field also changes as well as its magnitude in general.
We therefore need to attach, to each track segment, a local frame having its $z$ axis pointing to the magnetic field direction so as to use our helix parametrization defined in Eq.(\ref{eq:helixequation}).
At the end of each step, we hence update the frame to make its $z$ axis parallel with the magnetic field there and transform the propagated state vector to this new updated frame.

\begin{figure}[!htb]
\centering
\includegraphics{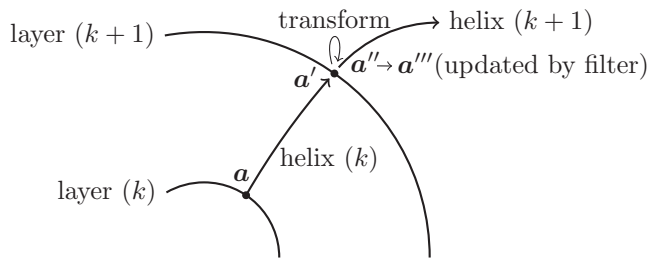}
\caption{Basic idea of transforming the state vector and the associated frame between two nearby layers.}
\label{fig::concept}
\end{figure}

This new track propagation procedure is illustrated in Fig.\ref{fig::concept}.
Firstly, the state vector $\boldsymbol{a}=\boldsymbol{a}_{k}$ at layer ($k$) is propagated to $\boldsymbol{a}^{\prime}$ at layer ($k+1$) along helix $(k)$ parametrized by Eq.(\ref{eq:helixequation}). 
The magnetic field at the intersection of the helix and layer ($k+1$) is then calculated. 
With the new magnetic field and the one at the last intersection, the matrix for rotating the frame can be obtained. 
Since the state vector is not easy to transform directly, it is converted to the momentum vector. 
Applying the rotation matrix to the momentum vector and then using the rotated momentum, we can calculate the predicted state vector $\boldsymbol{a}^{\prime\prime}$ in the new frame.
Notice that this defines a new modified state propagator as $\boldsymbol{a}^{\prime\prime} \equiv \boldsymbol{a}_{k+1}^{k} = \boldsymbol{f}^{\rm mod}_{k}(\boldsymbol{a}_{k})$.
Finally, we update the predicted state vector in the new frame with Eq.(\ref{eq::filter}), getting the filtered state vector $\boldsymbol{a}^{\prime \prime \prime} \equiv \boldsymbol{a}_{k+1}$. 
The same procedure is repeated for subsequent steps.

Since the propagation procedure or equivalently the propagator function is modified, the propagator matrix 
should also be modified accordingly.  We will explain how in the next subsection.

\subsection{Transformation}
\begin{figure}[!htb]
\centering
\includegraphics[width=8cm]{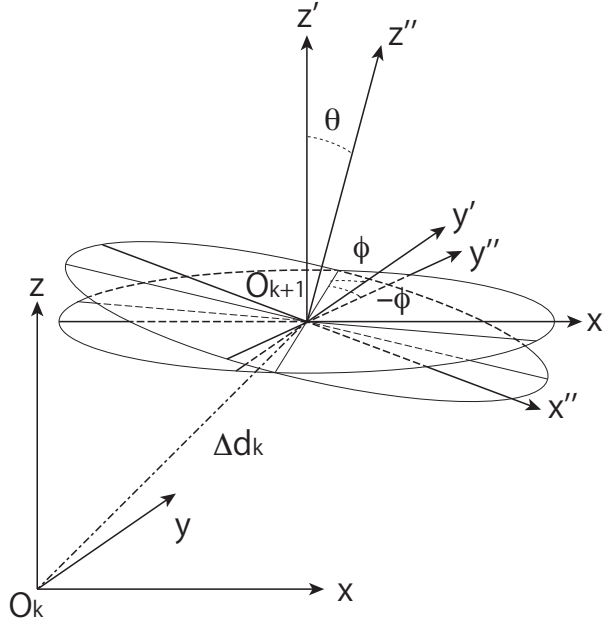}
\caption{Transformation from one frame to the next. The $\theta$ and $\phi$ angles are
determined by the magnetic field directions at the position $\boldsymbol{O}_{k}$ and 
$\boldsymbol{O}_{k+1}$.}\label{fig::transformation}
\end{figure}

The transformation between the two frames consists of a shift and a rotation as  is illustrated in Fig.\ref{fig::transformation}.
The shift is given by the vector $\Delta\boldsymbol{d}_{k}$ and parallel-transports the $xyz$-frame at the starting point on layer ($k$) to the new $x'y'z'$-frame at the predicted intersection on layer ($k+1$).
The $x'y'z'$-frame is then rotated so as to make the $z^{\prime \prime}$ axis point to the new magnetic field. 
This rotation $\boldsymbol{\Delta R}$ is defined as follows:
\begin{equation}\label{eq:rotmulti}
\boldsymbol{\Delta R}=
\boldsymbol{\Delta R}_{z^{\prime \prime}}(-\phi)
\boldsymbol{\Delta R}_{y^{\prime \prime}}(\theta)
\boldsymbol{\Delta R}_{z^{\prime}}(\phi),
\end{equation}
where the $x'y'z'$-frame is first rotated around the $z'$ axis by the angle $\phi$ to bring the $y'$ axis to a tentative $y''$ axis so that it becomes perpendicular to the plane spanned by the $z'$ axis and the new magnetic field direction; the frame is then rotated by the angle $\theta$ around the tentative $y''$ axis to make the $z''$  axis point to the new magnetic field direction; finally the frame is rotated again this time around the $z''$ axis by the angle $-\phi$ to bring the tentative $y''$ axis to its final direction.
%
%

Since the rotation is passive for a vector, the rotation matrices are given by
\begin{displaymath}
\boldsymbol{\Delta R}_{z^{\prime}}(\phi)=
\left(
\begin{array}{ccc}
\cos\phi & \sin\phi & 0\\
-\sin\phi & \cos\phi & 0\\
0 & 0 & 1
\end{array}
\right)
\end{displaymath}
and
\begin{displaymath}
\boldsymbol{\Delta R}_{y^{\prime \prime}}(\theta)=
\left(
\begin{array}{ccc}
\cos\theta & 0 & -\sin\theta\\
0 & 1 & 0\\
\sin\theta & 0 & \cos\theta
\end{array}
\right).
\end{displaymath}

Considering the shift and the rotation, a local position vector $\boldsymbol{x}_{k}$ in frame ($k$) can be transformed to a local position vector in frame ($k+1$) by
\begin{equation}\label{eq::localtransform}
\boldsymbol{x}_{k+1} = \boldsymbol{\Delta R}_{k}
(\boldsymbol{x}_{k} - \boldsymbol{\Delta d}_{k}).
\end{equation}
We also need  a transformation of a vector from a global frame to the local frame:
\begin{equation}\label{eq:globaltransform}
\boldsymbol{x}_{k+1} = \boldsymbol{R_{k}}( \tilde{\boldsymbol{x}} - \boldsymbol{d}_{k}),
\end{equation}
in which $\tilde{\boldsymbol{x}}$ is the corresponding vector defined in the global frame.
In Eq.(\ref{eq::localtransform}), $\boldsymbol{x}_k$ is defined in local frame ($k$), and it can be transformed from the global vector by
\begin{equation}\label{eq::transformcoord}
\boldsymbol{x}_{k} = \boldsymbol{R}_{k-1}( \tilde{\boldsymbol{x}} - \boldsymbol{d}_{k-1}) 
\end{equation}
Substituting Eq.(\ref{eq::transformcoord}) into Eq.(\ref{eq::localtransform})
\begin{displaymath}
\boldsymbol{x}_{k+1} = \boldsymbol{\Delta R}_{k}[ %
\boldsymbol{R}_{k-1}( \tilde{\boldsymbol{x}} - \boldsymbol{d}_{k-1}) %
- \boldsymbol{\Delta d_{k}} ], %
\end{displaymath}
then the global rotation matrix $\boldsymbol{R}_k$ and the shift vector $\boldsymbol{d}_k$ can be represented by the following recurrence formulae:
\begin{equation}
\left \{
\begin{array}{rll}
\boldsymbol{R}_{k} & = & \Delta\boldsymbol{R}_{k}\boldsymbol{R}_{k-1} \\
\boldsymbol{d}_{k} & = & \boldsymbol{d}_{k-1}+ \boldsymbol{R}_{k-1}^{-1} \Delta \boldsymbol{d}_{k} 
\end{array}.
\right .
\end{equation}

The local magnetic field in frame ($k$) is transformed simply by 
\begin{equation}
\boldsymbol{B} (\boldsymbol{x}) = \boldsymbol{R}_{k} \tilde{\boldsymbol{B}}
(\tilde{\boldsymbol{x}}),
\end{equation}
in which $\tilde{\boldsymbol{B}}(\tilde{\boldsymbol{x}})$ is the magnetic field in the global frame.

\subsection{Propagator matrix}
The total propagation described in Section \ref{subsec::basicidea} can be represented mathematically by
\begin{equation}\label{eq:vecarray}
\left \{
\begin{array}{lll}
\boldsymbol{a}^{\prime} &= &  \boldsymbol{f}_{k}(\boldsymbol{a}_{k})\\
\boldsymbol{p} & = & \boldsymbol{c}(\boldsymbol{a}^{\prime})\\
\boldsymbol{p}^{\prime} & = & \boldsymbol{t}( \boldsymbol{p})\\
\boldsymbol{a}^{\prime \prime} & = & \boldsymbol{c}^{-1}(\boldsymbol{p}^{\prime})
\end{array}
\right .,
\end{equation}
where, the function $\boldsymbol{f}_k$ is the original state vector propagation function for a uniform magnetic field, $\boldsymbol{c}$ is a function which converts a state vector to the corresponding momentum with $\boldsymbol{c}^{-1}$ being its inverse, and the function $\boldsymbol{t}$ is a rotation given by the rotation matrix $\boldsymbol{\Delta R}$. 

The momentum can be calculated using Eq.(\ref{eq:helixequation}):
\begin{equation}\label{eq::momentum}
\boldsymbol{p}=
-\left(\frac{Q}{\alpha}\right) \frac{d\boldsymbol{x}}{d\phi} 
= \frac{1}{|\kappa|}
\left(
\begin{array}{c}
-\sin\phi_0\\
\cos\phi_0\\
\tan\lambda\\
\end{array}
\right).
\end{equation}
As mentioned before, the sign of $\kappa$ is the sign of the particle charge. 
The momentum vectors in the two successive local frames are related by $\Delta \boldsymbol{R}$:
\begin{equation}\label{eq::momtrans}
\boldsymbol{p}^{\prime} = \Delta \boldsymbol{R} \, \boldsymbol{p} . 
\end{equation}
During the transformation, the intersection (taken to be the pivot) is on the helix. 
Therefore the two distance components, $d_{\rho}$ and $d_{z}$, are zero, 
while the other three components, $\phi_0$, $\kappa$, and $\tan\lambda$, can easily be solved from the momentum by their definitions:
\begin{equation}\label{eq::app}
\boldsymbol{a}^{\prime \prime}=
\left(
\begin{array}{c}
d_{\rho}\\
\mathrm{atan2} (-p_x^{\prime}, p_y^{\prime}) \\
\frac{s_{\kappa}}{\left(p_x^{\prime 2}+p_y^{\prime 2}\right)^{\frac{1}{2}}}\\
d_z\\
\frac{p_z^{\prime}}{\left(p_x^{\prime 2}+p_y^{\prime 2}\right)^{\frac{1}{2}}}\\
\end{array}
\right)  ,
\end{equation} 
where the sign of $\kappa$ in the last local frame (i.e. $s_{\kappa} \equiv \mathrm{sgn}(\kappa)$) is used, since
it is safe to assume that the magnetic field direction will not be reversed as long as the magnetic field varies moderately.

According to Eq.(\ref{eq:vecarray}), the modified propagator matrix is given by
\begin{equation}
\boldsymbol{F}_{k}^{\rm mod} \equiv
\frac{\partial \boldsymbol{f}^{\rm mod}_{k}(\boldsymbol{a})}{\partial \boldsymbol{a}}
=
\frac{\partial \boldsymbol{a}^{\prime \prime}}{\partial \boldsymbol{a}},
\end{equation}
which is calculated to be
\begin{equation}\label{eq:modprop}
\boldsymbol{F}_{k}^{\rm mod} =
\frac{\partial \boldsymbol{a}^{\prime \prime}}{\partial \boldsymbol{p}^{\prime}}
\frac{\partial \boldsymbol{p}^{\prime}}{\partial \boldsymbol{p}}
\frac{\partial \boldsymbol{p}}{\partial \boldsymbol{a}^{\prime}}
\frac{\partial \boldsymbol{a}^{\prime}}{\partial \boldsymbol{a}}
=\boldsymbol{F}_{k}^{\rm rot} \boldsymbol{F}_{k}.
\end{equation}
The original propagator matrix $\boldsymbol{F}_{k}$ is known in the original KalTest already. 
The other three matrices in $\boldsymbol{F}_{k}^{\rm rot}$ are given in \ref{sec::appendixmatrix}.


The transformation algorithm described in this section is implemented in the class {\tt TTrackFrame}. 
In order to promote our {\tt THelicalTrack} class to accommodate the segment-wise helical track model for non-uniform magnetic field, a {\tt TTrackFrame} object is added as its data member. 
With the help of {\tt TTrackFrame}, the {\tt THelicalTrack} can calculate the global position, the modified 
state vector, and the modified propagator matrix.
This way, all the modifications for the segment-wise helical track implementation are hidden in {\tt THelicalTrack} and
are invisible to the other classes.
Because of the locality of the segment-wise helical track model implementation, it can be seen that the track smoothing is still valid from Eq.(\ref{EQ:Ak}). 
Further it should be relatively easily ported into any Kalman filter package other than KalTest.

\section{Test of algorithm}\label{sec::results}
\subsection{Simulation conditions}
After the algorithm for track fitting in a non-uniform magnetic field was implemented based on the original KalTest code, its track reconstruction performance has been tested intensively. 
Firstly, the  mean and the sigma of momentum distributions were compared for the original code and the new code. 
As expected, the results are identical if tracks are generated and reconstructed in a uniform magnetic field.

For the purpose of testing track reconstruction in a non-uniform magnetic field, we assumed the following (artificially non-uniform) magnetic field $\boldsymbol{B}$:
\begin{displaymath}
\left \{
\begin{array}{rllccc}
B_x & = & B_0 k x z \\
B_y & = & B_0 k y z \\
B_z & = & B_0 (1 - k z^2)
\end{array}
\right.,
\end{displaymath}
in which, $k=\frac{k_0}{z_{\rm max} r_{\rm max}}$, $B_0 = 3  \ \mathrm{T}$, $z_{\rm max} = r_{\rm max} = 3000 \ \mathrm{mm}$. 
The coefficient $k$ can be set to several values for different non-uniformities. 
In simulation, we assumed a tracker having 251 measurement layers with a layer-to-layer distance of $6 \ \mathrm{mm}$, an inner radius of $300 \ \mathrm{mm}$, and outer radius of $1800 \ \mathrm{mm}$. 
The detector geometry is similar to the configuration of the LC TPC in the ILD detector concept\cite{Abe:2010aa}. 

\begin{figure}[!htb]
\centering
\setcounter{subfigure}{0}%
  \subfigure[$xy$ view]{
   \includegraphics[scale=0.3]{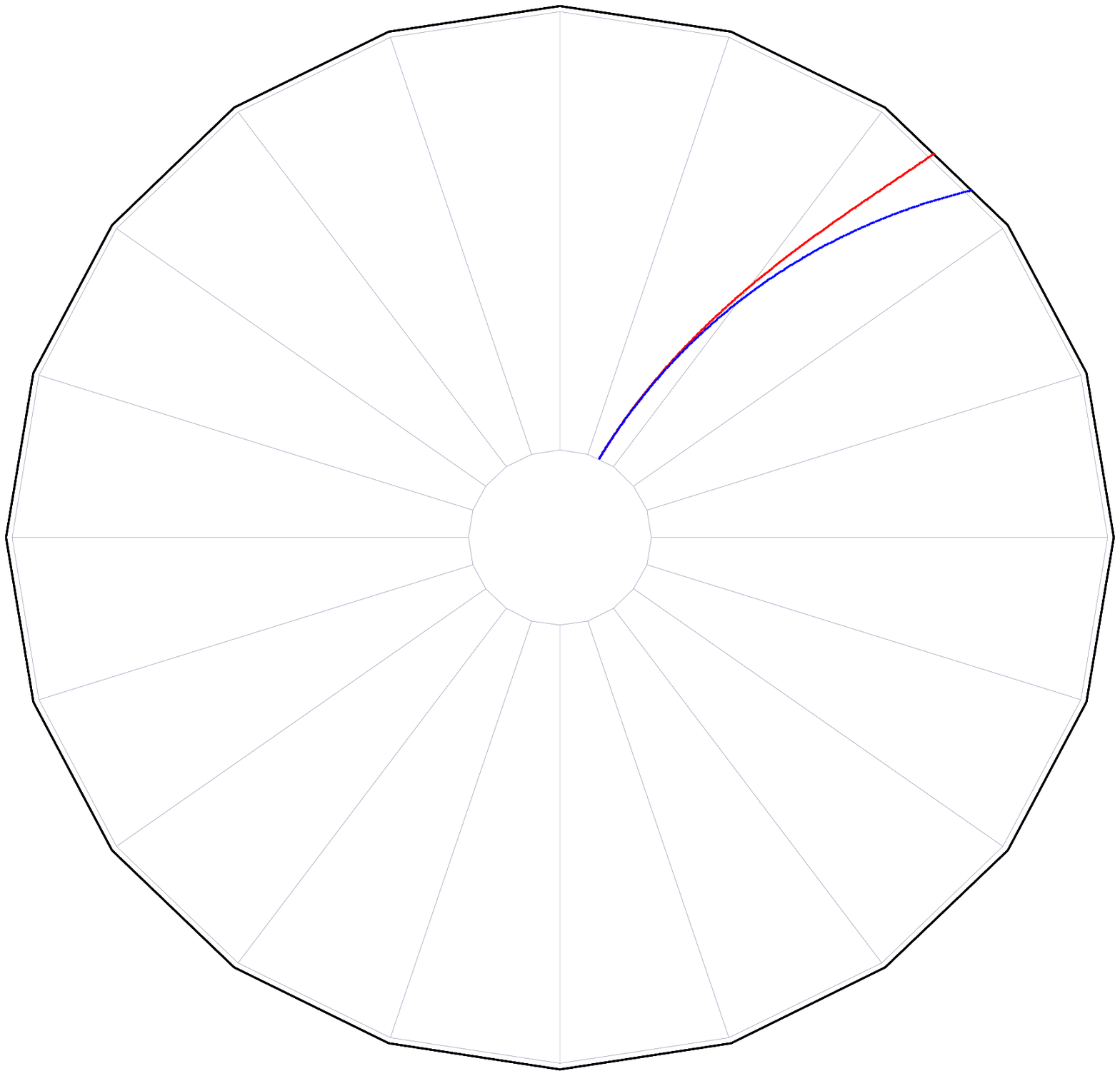}\label{fig::track2_xy}
  }
  \subfigure[3D view]{ 
  \includegraphics[scale=0.3]{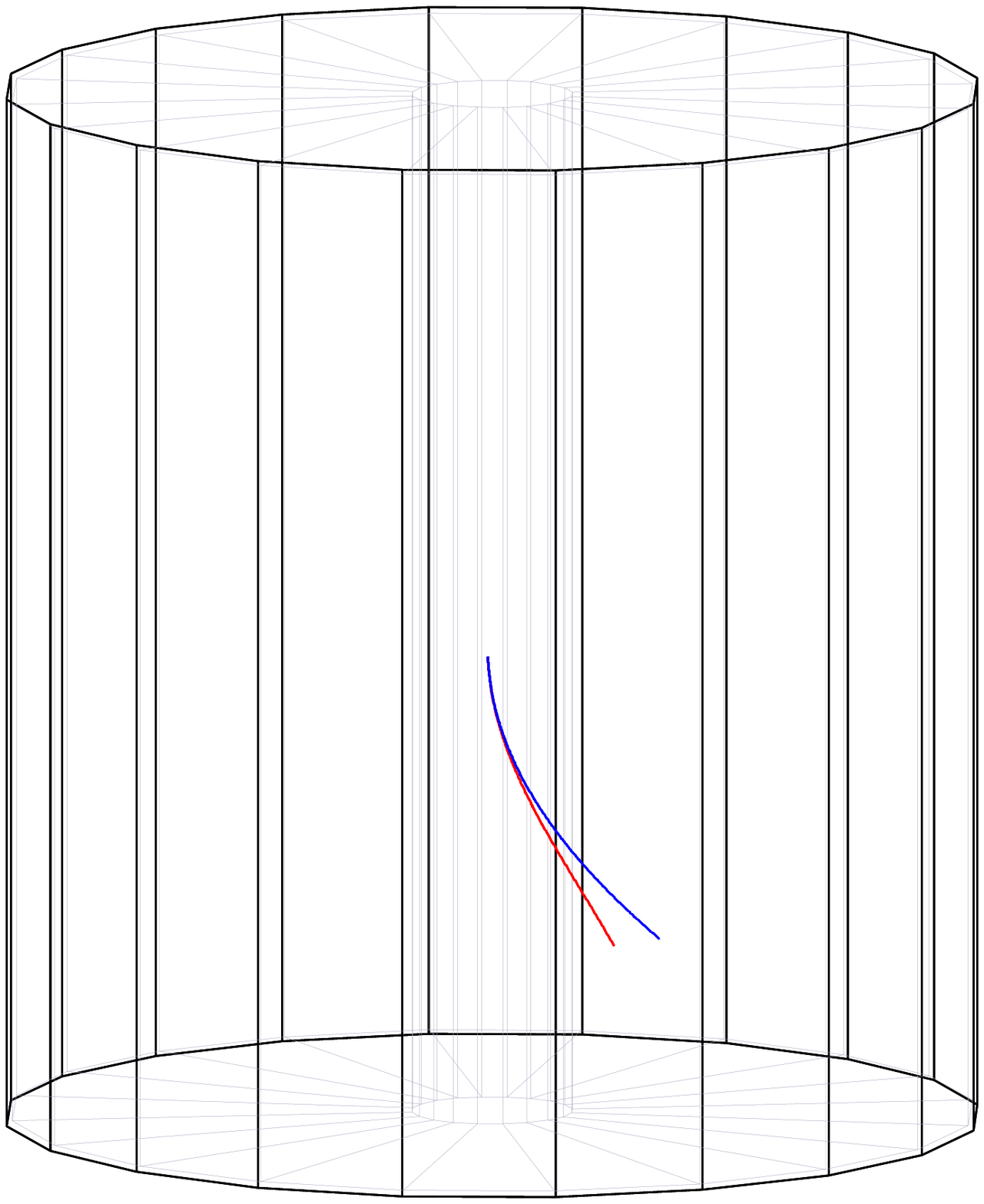}\label{fig::track2_3d}
  }
  \caption{Event display. Tracks with $2 \ \mathrm{GeV/}c$ simulated in a non-uniform magnetic field (red curve: $k_0 = 5$) and uniform magnetic field (blue curve: $k_0=0$).}
  \label{fig::eventdisplay}
\end{figure}

To simulate tracks in the non-uniform magnetic field, a Runge-Kutta propagator class in ROOT\cite{Antcheva:2011zz}, {\tt TEveTrackPropagator}, was used. 
In this class, the position is a function of the length along the track and the crossing point of the track with a measurement layer is calculated by the bisection method. 
The crossing point was smeared to make a hit position according to the detector spatial resolution ($100 \ \mathrm{\upmu m}$) in the $r \phi$ plane. 
For the track generation, we generated dip angles and an azimuthal angles uniformly in $\lambda \in [0, 0.5]$ and  $\phi \in [0, 2\pi]$, respectively.
In Fig.\ref{fig::eventdisplay}, a typical event display is shown for a track generated in the non-uniform magnetic field with $k_0=5$.
In the same figure, for comparison, a track with the same initial momentum but in a uniform magnetic field ($k_0=0$) is also shown. 
Notice that the third component of $\boldsymbol{B}$ depends on the $z$ position quadratically in our field assumption, therefore the discrepancy of the two tracks increases as their $z$ position becomes large. 
%

\subsection{Results}
\subsubsection{Momentum measurement}
\begin{figure}
\centering
\setcounter{subfigure}{0}%
  \subfigure[Reconstructed in uniform magnetic field]{
   \includegraphics[scale=0.7]{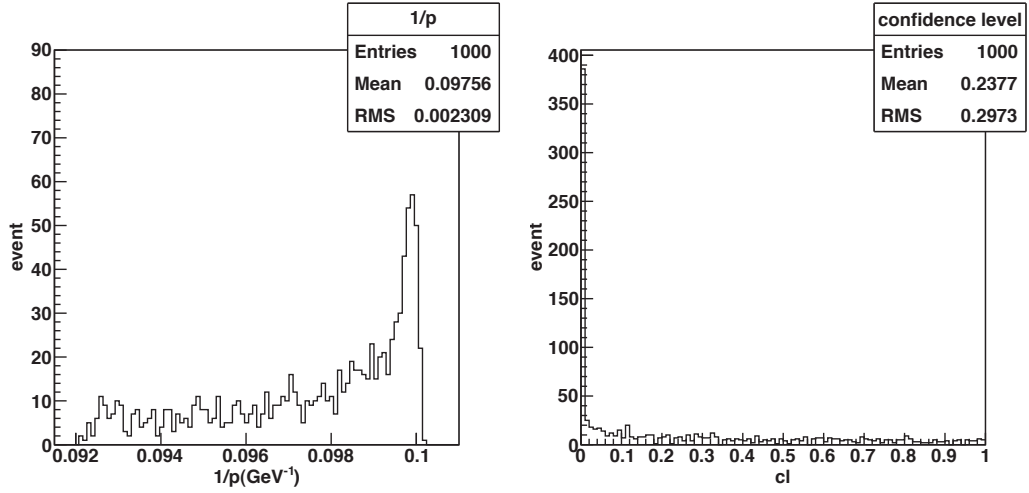}\label{fig::resultsuni}
  }
  \\
  \subfigure[Reconstructed in non-uniform magnetic field]{ 
  \includegraphics[scale=0.7]{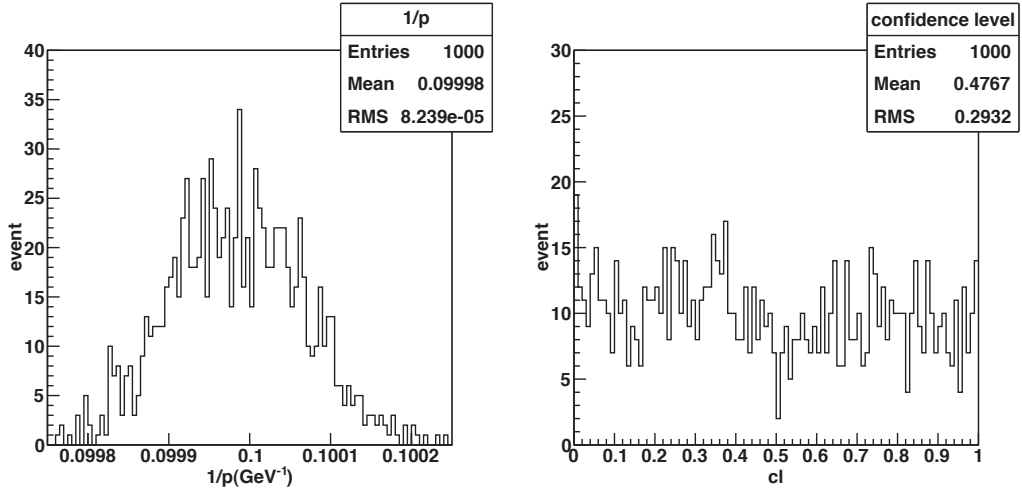}\label{fig::resultsnon}
  }
  \caption{Comparison of track-fit results with the original helical track model and those with the segment-wise helical track model for tracks generated in a non-uniform magnetic field. 
The tracks were generated with the coefficient $k_0$ of the non-uniformity being 1, and the momentum $10 \ \mathrm{GeV/}c$.}
  \label{fig::comparison}
\end{figure}

With tracks generated in the non-uniform magnetic field, we compared the reconstructed reciprocal momentum and confidence level distributions for the track-fit results using the original helical track model assuming a uniform field and those using the segment-wise helical track model in Fig.\ref{fig::comparison}.
The total momentum is calculated as $p=p_t \sqrt{1+\tan^2 \lambda}$, in which $p_t$ and $\tan \lambda$ are the fitted parameters in the state vector. 
In Fig.\ref{fig::resultsuni}, we used the original helical track model with a uniform magnetic field: $\boldsymbol{B} = (0, 0, 3 \ \mathrm{T})$ to reconstruct the tracks. 
The distribution of the reciprocal momentum on the left panel of Fig.\ref{fig::resultsuni} is non-gaussian, and the confidence level distribution on the right panel has a delta-function-like peak at zero, indicating that the fitting is inconsistent with the simulated tracks. 
This is exactly why the simple helical track model used in the original KalTest must be updated for the non-uniform magnetic field situation for future linear collider experiments.
%

The result reconstructed with the segment-wise helical track model assuming the same magnetic field as that of the track generation is shown in Fig.\ref{fig::resultsnon}. 
The mean value of the momentum distribution on the left panel is very close to the expected, $0.1 \ \mathrm{(GeV/}c)^{-1}$, and its standard deviation is consistent with Gluckstern's formula\cite{Gluckstern:1963ng}. 
The flat confidence level distribution on the right panel implies the fitted tracks and hits are also consistent. 
Therefore, from Fig.\ref{fig::resultsuni} and \ref{fig::resultsnon}, it can be concluded that the improved algorithm with the segment-wise helical track model performs much better in track reconstruction since the non-uniformity of the magnetic field is properly taken into account. 

In fact, the bad resolution in Fig.\ref{fig::resultsuni} was found to come from a momentum bias. 
This is because the magnetic field becomes weaker as $z$ increases, although it is assumed to be constant in the reconstruction with the original helical track model, thereby overestimating $p_t$.
Since the magnetic field decreases with the $z$ coordinate, as the dip angle increases, the reconstructed $1/p_t$ decreases. 

The results for different non-uniformities and dip angles with a track step size (i.e. helix segment size) of 6 $\mathrm{mm}$ are given in Table \ref{table::rms_stepsize6mm}. 
As expected, for a specified dip angle, the momentum bias becomes more prominent as the non-uniformity increases, since the assumption that the magnetic field is approximately uniform between two nearby layers becomes inappropriate.  
By the same token, the bias grows with the dip angle. 
Table \ref{table::rms_stepsize1mm} tells us that the mean value can be improved by adding dummy stepping layers to make the helix segment short enough and consequently the algorithm is forced to reevaluate the magnetic field at more stepping points. 
Tables \ref{table::rms_stepsize6mm} and \ref{table::rms_stepsize1mm} for the different step sizes have consistent momentum resolutions and as the non-uniformity becomes large the momentum resolution also increases because the average magnetic field experienced by the tracks becomes smaller. 
This explanation is also valid for the dip angle dependence for a fixed $k_0$ value (2 or 3). 
For the low non-uniformity values (0.5 or 1), the momentum resolution is dominated by the lever arm length and hence 
as $\lambda$ increases, the transverse momentum gets smaller, and the resolution gets better.
%
\begin{table}
  \caption{Mean and RMS of $1/p$ (in units of $10^{-1} \cdot \mbox{(GeV/}c)^{-1}$ and $10^{-5} \cdot \mbox{(GeV/}c)^{-1}$, respectively)}
  \label{table::rms}
  \centering
  \subtable[Step size $6 \ \mathrm{mm}$]{
  \begin{tabular}{ccccccccc}
  \toprule
      {$k_0$} & {$\lambda = 0.1$} &  {$\lambda = 0.3$}  & {$\lambda = 0.5$}
        \\\midrule
       0.5 & 1.0000/8.02 & 0.9999/7.79 & 0.9998/7.34\\
      1     & 1.0000/8.03 & 0.9998/7.89 & 0.9995/7.65\\
      2     & 1.0000/8.05 & 0.9997/8.09 & 0.9990/8.36\\
      3     & 0.9999/8.07 & 0.9995/8.31 & 0.9984/9.20\\\bottomrule
  \end{tabular}
  \label{table::rms_stepsize6mm}
  }
  \qquad
  \subtable[Step size $1 \ \mathrm{mm}$]{
  \begin{tabular}{ccccccccc}
  \toprule
      {$k_0$} & {$\lambda = 0.1$} &  {$\lambda = 0.3$}  & {$\lambda = 0.5$}
      \\\midrule
      0.5 & 1.0000/8.02 & 1.0000/7.79 & 1.0000/7.34\\
      1    & 1.0000/8.03 & 1.0000/7.89 & 0.9999/7.65\\
      2    & 1.0000/8.05 & 0.9999/8.10 & 0.9998/8.36\\
      3    & 1.0000/8.07 & 0.9999/8.32 & 0.9997/9.21\\\bottomrule
  \end{tabular}
  \label{table::rms_stepsize1mm}
  }
\end{table}

\subsubsection{CPU time}
The time consumption of relevant functions in the track fitting (without the dummy layers to force the stepping size to be $1\,\mathrm{mm}$) for 1000 tracks is listed in Table \ref{table::time}. 
These CPU times were measured on a laptop PC with a $2.4\,\mathrm{GHz}$ Intel Core 2 Duo processor.
The sum of CPU times consumed by track propagation and filtering turned out to essentially be the total track fitting time. 
From Table \ref{table::time}, one can see that the time consumption increased by about 10 seconds because of using the track frame transformation class, which was absent in the original KalTest.
This means that the CPU expense was approximately doubled by the introduction of the segment-wise helical track model.
The functions transforming vectors are called many times mostly in the function calculating the crossing points. 
One possibility to reduce this time consumption is to optimize the number of calls for the crossing point calculating function.
\begin{table}
  \caption{Time expense of relevant functions with the segment-wise helical track model for 1000 tracks in units of $\mathrm{sec.}$.}
  \label{table::time}
  \centering
  \begin{tabular}{llccccccc}
  \toprule
      {Function} & {Time expense}
        \\\midrule
      Total & 18.82\\
      TVKalState::Propagate & 11.53\\
      TVKalSite::Filter & 7.27\\ 
      TTrackFrame::TTrackFrame & 0.87 \\
      TTrackFrame::TransformVector & 6.59\\
      TTrackFrame::TransformSv & 2.58\\
       TVSurface::CalcXingPointWith  & 5.90\\\bottomrule
  \end{tabular}
\end{table}

\section{Summary}
The helical track model in KalTest has been updated for the track fitting in a moderately non-uniform magnetic field by introducing the concept of segment-wise helical track model and implemented in the {\tt THelicalTrack} class.
The segment-wise helical track model assumes that the magnetic field is approximately uniform between two nearby measurement sites and approximates the track segment between the two sites to be a helix determined by the track momentum and the magnetic field at the starting site of the step.
The non-uniformity of the magnetic field is then taken into account by reevaluating the magnetic field at the end site of the step for the next step.
This requires the segment-wise helical track object {\tt THelicalTrack} to carry a coordinate frame object {\tt TTrackFrame} as its data member to specify the local coordinate system in which the state vector consisting of the helix parameters is defined.
The helix frame is then transformed at each step so as to make the local $z$ axis pointing to the new magnetic field direction.
The coordinate transformation modifies the original state propagator and its derivative.
In this paper we have elaborated the mathematical formulation of the modifications for Kalman-filter-based tracking.
It should be emphasized that the modifications are localized in the {\tt THelicalTrack} class and hence are invisible to the other part of the KalTest classes.
This significantly facilitated the implementation of the code to handle the non-uniform magnetic field since there was essentially no need to touch the original KalTest architecture except for the track model {\tt THelicalTrack} and its supporting class {\tt TTrackFrame}.
The algorithm should hence be relatively easy to port to any other Kalman-filter-based track fitting package.

We tested the segment-wise helical track model and its performance in Kalman-filter-based track fitting in a non-uniform magnetic field.
The test showed that the track fitting with the segment-wise helical track model works very well for a modest field non-uniformity and yields correct track momentum values in the non-uniform magnetic field. 
It was also demonstrated that the track fitting performance can be enhanced by adding dummy layers (stepping layers) to reduce the step size so that the track fitting works in a highly non-uniform field situation. 

The CPU time expense was measured and found to be approximately doubled as compared to the original KalTest for a uniform magnetic field.
The increase was mostly due to the repeated coordinate transformations for various objects in stepping, calculations of crossing points of the track and measurement layers in particular.  
It can probably be improved by optimizing the code for the crossing point calculations.

The source code of the new KalTest with the segment-wise helical track model can be downloaded from \url{http://www-jlc.kek.jp/jlc/en/subg/soft/tracking/kaltest-nonuniform}.
\section*{Acknowledgment}
The original KalTest package was developed in the frame work of the LCTPC collaboration. We would like to thank all the developers including K.\,Hoshina, Y.\,Nakashima, and A.\,Yamaguchi for their significant contributions in the early stage of the development.
The authors are grateful to the ILD software team and the members of the LCTPC group for useful discussions and supports for this work. Among them Steve Aplin and Frank Gaede deserve special mention. Their efforts to integrate the original KalTest package into the standard ILD event reconstruction chain and to test it throughly in various simulation studies set a firm basis on which this work could be built. The first author would like to thank the support from ILC group of Institute of Particle and Nuclear Studies, KEK and LCD group of CERN. He also thanks to useful advice from Martin Killenberg.
This work is supported in part by the Creative Scientific Research Grant No. 18GS0202 of the Japan Society for Promotions of Science (JSPS), the JSPS Core University Program, and the JSPS Grant-in-Aid for Science Research No. 22244031, and the JSPS Specially Promoted Research No. 23000002. This work is also supported in part by National Natural Science Foundation of China under Contract Nos. 11075084, 11161140590.



\appendix

\section{The modified propagator matrix}\label{sec::appendixmatrix}
The concrete form of $\boldsymbol{F}^{\rm rot}_{k}$ in Eq.(\ref{eq:modprop}) is shown here.  
Notice first that at each step before the transformation to the new frame, the pivot of the predicted state vector is temporarily taken to be the predicted intersection of the track with the measurement
layer there, implying that $d_\rho$ and $d_z$ are zero.\footnote{After the frame transformation the 
pivot is moved to the actual hit position from the predicted intersection.}
According to Eq.(\ref{eq::momentum}), the momentum calculation needs only the three non-zero parameters of the state vector. 
Now let the state vector $\boldsymbol{a}' =
(\phi_{0},\kappa,\tan\lambda)^{\mathrm{T}}$, then the calculation of the first of the three derivatives in $\boldsymbol{F}_{k}^{\rm rot}$ is straightforward:
\begin{equation}
\frac{\partial \boldsymbol{p}}{\partial \boldsymbol{a}^{\prime}}=
\left(
\begin{array}{ccccc}
-\frac{1}{|\kappa|}\cos\phi_0 & \frac{s_{\kappa}}{{\kappa}^2}\sin\phi_0 & 0 \\
-\frac{1}{|\kappa|}\sin\phi_0 & -\frac{s_{\kappa}}{{\kappa}^2}\cos\phi_0  & 0 \\
0 & -\frac{s_{\kappa}}{{\kappa}^2}\tan\lambda & \frac{1}{|\kappa|}
\end{array}
\right).
\end{equation}
According to Eq.(\ref{eq::momtrans}), the second of the three derivatives in $\boldsymbol{F}_{k}^{\rm rot}$ is
\begin{equation}
\frac{\partial \boldsymbol{p}^{\prime}}{\partial \boldsymbol{p}}=
\Delta \boldsymbol{R}.
\end{equation}
If we just use the non-zero components in Eq.(\ref{eq::app}), the derivative of the new state vector with respect to the momentum is
\begin{equation}
\frac{\partial \boldsymbol{a}^{\prime \prime}}{\partial \boldsymbol{p^{\prime}}}=
\left(
\begin{array}{ccccc}
-\frac{p_y}{p_{\mathrm{T}}^2} & \frac{p_x}{p_{\mathrm{T}}^2} & 0 \\
-\frac{s_{\kappa} p_x}{p_{\mathrm{T}}^3} & 
-\frac{s_{\kappa} p_y}{p_{\mathrm{T}}^3}& 0 \\
-\frac{p_x p_z}{p_{\mathrm{T}}^3} & -\frac{p_y p_z}{p_{\mathrm{T}}^3} & \frac{1}{p_{\mathrm{T}}^3}
\end{array}
\right).
\end{equation}
The product of the three matrices:
\begin{displaymath}
\boldsymbol{M} = \frac{\partial \boldsymbol{a}^{\prime \prime}}{\partial \boldsymbol{p^{\prime}}}%
\frac{\partial \boldsymbol{p}^{\prime}}{\partial \boldsymbol{p}} %
\frac{\partial \boldsymbol{p}}{\partial \boldsymbol{a}^{\prime}}
\end{displaymath}
is a $3 \times 3 $ matrix. 
To use it in Eq.(\ref{eq:modprop}), the elements with zero values should be restored, namely
\begin{equation}
\boldsymbol{F}_{k}^{\rm rot} = \left(
\begin{array}{cccccc}
 1 & 0            & 0           & 0 & 0            \\
 0 & M_{00} & M_{01} & 0 & M_{02} \\
 0 & M_{10} & M_{11} & 0 & M_{12} \\
 0 & 0            & 0            & 1 & 0            \\
 0 & M_{20} & M_{21} & 0 & M_{22}
\end{array}
\right).
\end{equation}



\bibliographystyle{elsarticle-num}
\bibliography{refs}







\end{document}